\documentclass[aps,prl,twocolumn,showpacs,groupedaddress]{revtex4-1}
\usepackage{graphicx}  
\usepackage{dcolumn}   
\usepackage{bm}        
\usepackage{amssymb}   
\usepackage{amsmath}
\usepackage{braket}
\usepackage[mathscr]{euscript}
\begin{document}
\title{High-Fidelity Entangled Bell States via Shortcuts to Adiabaticity}
\author{Koushik Paul}
\author{Amarendra K Sarma}
\email{aksarma@iitg.ernet.in}
\affiliation{Department of Physics, Indian Institute of Technology Guwahati\\ Guwahati-781039, Assam, India}
\date{\today}
\begin{abstract}
We present a couple of protocols based on shortcut to adiabaticity techniques for rapid  generation of robust entangled Bell states in a system of two two-state systems. Our protocols rely on the so-called transitionless quantum driving (TQD) algorithm and Lewis-Riesenfeld invariant (LRI) method. Both TQD and LRI methods result in high fidelity in population transfer.Our study shows that it is possible to prepare an entangled state in infinitely short time without losing robustness and efficiency. 
\end{abstract}
\maketitle
{
In the realm of quantum physics, adiabatic passage protocols are most widely and rigorously exploited in last few decades for various applications related to coherent control and quantum information processing \cite{1}. The simplicity and robustness of this method leads to numerous applications in various systems across different branches of physics, such as CTAP (Coherent Tunneling by Adiabatic Passage) in quantum dots \cite{2},adiabatic directional couplers \cite{3}, superconducting Josephson junctions \cite{4}, Bose-Einstein condensates \cite{5}, population manipulation among atomic and molecular states \cite{6}, Transport in optical lattices \cite{7} etc. One of the  advantages of adiabatic passage technique is its robustness towards atomic decoherence on the fidelity.  However the main shortcoming in this method is that it follows adiabatic evolution that makes it sufficiently slow with respect to time. The quest to overcome this limitation ends up in a set of new theories that are nowadays known as the shortcut techniques. Among them, two methods gained particular attention recently, namely {\it Transitionless Quantum Driving} (TQD) or {\it Counterdiabatic algorithm} \cite{8,9} and {\it Lewis-Riesenfeld Invariant} (LRI) based approach \cite{10}. TQD permits us to control the time evolution of a particular system exactly along the adiabatic path beyond adiabatic limits by means of additional interactions. On the other hand, LRI based approach follows the invariant dynamics determined by appropriate boundary conditions regardless of adiabatic path and adiabatic conditions. According to these theories, quantum systems can be driven in an infinitely small amount of time without affecting the robustness of the process. Several extensive studies have been carried out in this direction in recent years \cite{11,12,13}.

Entanglement is another key concept in modern quantum physics. Preparation of different entangled states such as entangled atomic states, entangled states inside cavity and in different optical arrangements \cite{14} has remained a challenge in the context of quantum information and quantum optics. In fact few recent works show that adiabatic following can be implemented successfully for preparation of entangled states. In this context, the work by R. G. Unnayan et al. \cite{15}  is of tremendous significance, in which they described the adiabatic evolution from an un-entangled state to an entangled state. They considered a combination of two two-state systems, consists of two spin $\frac{1}{2} $ particles coupled by their intrinsic exchange interaction and an external time dependent magnetic field. Such a system, when magnetic field is switched off, can be represented in terms of conventional singlet and triplet states, among which two are basically the entangled Bell states for a bipartite system. With magnetic field being switched on, these four states get coupled to each other. Choosing these four states as basis, one can readily study the adiabatic evolution of this system. This work triggered a series of studies relating adiabatic passage and entangled state preparation afterwards \cite{16,17}. However the issue of long preparation time, inherent with the adiabatic passage method, still remains. In this work, our goal is to overcome this issue. Hence we study the possibilities for preparation of entangled states in a similar system by using above stated shortcut techniques.

{ \it Adiabatic method}- 
Let us consider two spin $\frac{1}{2} $ particles, $A$ and $B$, be coupled via exchange interaction and a time dependent magnetic field ${\mathbf B}(t)$ along the $z$ direction. The
Hamiltonian for such a system is given by \cite{15}:
\begin{eqnarray}
H(t) = 4\xi \hat{S}_A^z \otimes \hat{S}_B^z + \mu {\mathbf B}(t).(\hat{\mathbf S}_A + \hat{\mathbf S}_B)
\label{initial_ham}
\end{eqnarray}
Here $\xi$ denotes the exchange interaction parameter and $\mu$ is the gyromagnetic ratio. $\hat{\mathbf S}_A$ and $\hat{\mathbf S}_B$ are the respective spin operators. The first term of $H(t)$ is time independent ($H_0$) and its eigenstates are as follows:
\begin{subequations}
\begin{align}
\ket{\psi_{\uparrow \uparrow}} &= {\ket{\uparrow}}_A {\ket{\uparrow}}_B\\
\ket{\psi_{\downarrow\uparrow}^+} &= \frac{1}{\sqrt{2}}({\ket{\uparrow}}_A {\ket{\downarrow}}_B + {\ket{\downarrow}}_A {\ket{\uparrow}}_B)\\
\ket{\psi_{\downarrow\downarrow}} &= {\ket{\downarrow}}_A {\ket{\downarrow}}_B\\
\ket{\psi_{\downarrow\uparrow}^-} &= \frac{1}{\sqrt{2}}({\ket{\uparrow}}_A {\ket{\downarrow}}_B - {\ket{\downarrow}}_A {\ket{\uparrow}}_B),
\label{arrows}
\end{align}
\end{subequations}
where $\ket{\uparrow}$ and $\ket{\downarrow}$  stands for spin-up and spin-down states
respectively. ${\ket{\uparrow}}_A {\ket{\uparrow}}_B$ denotes the tensor (or direct) product,
${\ket{\uparrow}}_A \otimes {\ket{\uparrow}}_B$, between these two states. Upon choosing
those states as the basis and with the magnetic field being taken into account, the interaction Hamiltonian could be written as ($\hbar = 1$),
\begin{equation}
H_I(t) = \begin{pmatrix} \omega-\overline{B}_z(t) & \frac{1}{\sqrt{2}}\Omega(t) & 0\\
 \frac{1}{\sqrt{2}}\Omega(t) & -2\xi & \frac{1}{\sqrt{2}}\Omega(t))\\
0 & \frac{1}{\sqrt{2}}\Omega(t)) & -\omega+\overline{B}_z(t) \end{pmatrix}
\label{3level_ham}
\end{equation}
Since the magnetic field is time dependent and it contributes to both the diagonal and the off-diagonal terms of the interaction Hamiltonian, its choice is crucial. For our calculations we have chosen the magnetic field components as follows:
$ \overline{B}_x(t) = \Omega(t) \cos(\omega t)$, $ \overline{B}_y(t) = \Omega(t) \sin(\omega t)$,
$ \overline{B}_z(t) = \alpha^2 t$, where $\overline{\mathbf B}(t) = \mu {\mathbf B}(t)$ and $\alpha$ is a parameter
having the dimension of frequency. The diagonal elements in Eq.~(\ref{3level_ham}) are generally known as the diabatic energies. They represent the original base states (Eq.~(\ref{initial_ham})) in the interaction picture (therefore can be called diabatic states) and cross
each other at different times, $t_{12} = (\omega+2\xi)/\alpha^2$, $t_{13} = \omega/\alpha^2$ and $t_{23} = (\omega-2\xi)/\alpha^2$, producing the so called level crossings. However, in the context of adiabatic dynamics, the level crossing is avoided by choosing the external field
($\Omega(t)$) centered around the time where the crossing occurs. We choose $\Omega(t)$ to be centered at $t_{12}$ where the first two energies cross each other. To achieve the transition between the first two states it can be taken as,
\begin{equation}
\Omega(t) = \Omega_0 \exp[-(t-t_{12})^2/T^2]
\end{equation}
Since $\Omega(t)$ is centered at $t_{12}$, it effectively remains zero at other crossings. As a result the interaction is restricted only between the first two diabatic states \cite{15} and therefore we rewrite Eq.~(\ref{3level_ham}) as,
\begin{equation}
H_I(t) = \begin{pmatrix} \frac{\Delta(t)}{2} & \frac{1}{\sqrt{2}}\Omega(t) \\
 \frac{1}{\sqrt{2}}\Omega(t) & -\frac{\Delta(t)}{2} \end{pmatrix}
\label{2level_ham}
\end{equation}
Here $\Delta(t) = \omega + 2\xi - \alpha^2 t$. The adiabatic condition near $t_{12}$ is given by: $Q\ll1$, where $Q(t_{12}) = \alpha^2/2\Omega_0^2$. In Fig.~(\ref{fig1}a), we numerically plotted the population of the entangled state $\ket{\psi_{\downarrow\uparrow}^+}$ while varying $\alpha$ and $\Omega_0$. It clearly shows that for small $\Omega_0$ and larger $\alpha$, population of $\ket{\psi_{\downarrow\uparrow}^+}$ does not reach to maximum. Also a critical value of $\alpha = \alpha_c \sim 0.25\xi$ is required for population transfer as, for $\alpha$ being less than $\alpha_c$ does not allow $\Delta(t)$ to reach the crossing. For $Q=0.1$, the evolution of population from $\ket{\psi_{\uparrow \uparrow}}$ to $\ket{\psi_{\downarrow\uparrow}^+}$ is shown in Fig.~(\ref{fig1}a), where population increases adiabatically from 0 to 1 in the entangled state.

{\it Transitionless Driving}- The instantaneous (or adiabatic) eigenvectors of $H_I(t)$ are given by
\begin{equation}
[\ket{\phi_+(t)}, \ket{\phi_-(t)}]^T = U(\theta(t))^\dagger[\ket{\psi_{\uparrow \uparrow}}, \ket{\psi_{\downarrow\uparrow}^+}]^T
\label{adia_basis}
\end{equation}
$U(\theta(t))$ represents a 2D axis rotation where $\theta(t)$ is the angle of mixing and can be expressed as $\theta(t) = \tan^{-1}[-{\sqrt{2}}\Omega(t)/\Delta(t)]$. These states also satisfy the Schr{\"o}dinger equation and the interaction Hamiltonian can be expressed in the $\ket{\phi_{\pm}(t)}$ basis (adiabatic basis) by using a time dependent unitary transformation, $H_a(t) = U^{\dagger}H_I(t)U - iU^{\dagger}\dot{U}$.  $H_a(t)$ is generally known as the adiabatic Hamiltonian. According to Berry's algorithm of Transitionless quantum driving, it is always possible to construct a driving Hamiltonian, which cancels out the non-adiabatic part from the adiabatic Hamiltonian. Addition of a driving term in  $H_a(t)$ drives the system exactly along the adiabatic path even when adiabatic limit is crossed. The driving Hamiltonian, $H_1(t)$ is constructed from the instantaneous eigenstates which is Hermitian and purely off-diagonal in nature, can be written in adiabatic basis as \cite{8},
\begin{figure}
\includegraphics[height=8cm,width=8cm]{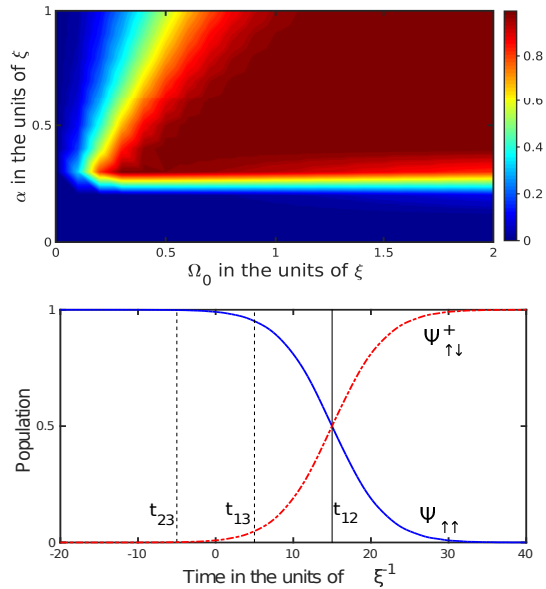}
\caption{\label{fig1} (Color online) (a) Final population of the entangled state $\ket{\psi_{\downarrow\uparrow}^+}$ against $\Omega_0$ and $\alpha$ with the parameters chosen as $\omega = \xi,\;T = 20\xi^{-1}$, (b) Evolution of population when adiabatic condition is
satisfied with $Q = 0.1$}
\end{figure}
\begin{multline}
H_1(t) = i\sum_m[\ket{\partial_t\phi_m(t)}\bra{\phi_m(t)}\\
- \braket{\phi_m(t)|\partial_t\phi_m(t)}\ket{\phi_m(t)}\bra{\phi_m(t)}]
\label{driving_ham}
\end{multline}
$H_1(t)$ can be realized by introducing another extra field to
the system which can be of different form in different systems. With the application of
$H_1(t)$, the adiabatic evolution will be followed in infinitely short time even with
smaller field amplitude $\Omega_0$ and rapid $\Delta(t)$ variation. Using Eq.~(\ref{adia_basis}) and Eq.~(\ref{driving_ham}), we calculate the driving Hamiltonian in
the diabatic basis as follows:
\begin{equation}
H_1(t) = \begin{pmatrix} 0 & i\Omega_a(t) \\ -i\Omega_a(t) & 0 \end{pmatrix}
\label{omega_a}
\end{equation}
Here $\Omega_a = \dot{\theta}/2$ represents the additional driving interaction. This can solely be evaluated from the mixing angle itself, which indeed makes the peak value of $\Omega_a(t)$ comparable with $\Omega_0$. The total Hamiltonian, in $[\ket{\psi_{\uparrow \uparrow}}, \ket{\psi_{\downarrow\uparrow}^+}]^T$ basis, to perform TQD will be $H_f(t) = H_I(t)+H_1(t)$. Addition of $H_1(t)$ includes an additional phase, $\zeta(t) = 2 {\tan^{-1}}(-\dot{\theta}(t)/\sqrt{2}\Omega(t))$. To simplify things further, another unitary transformation can be introduced. This leads to:
\begin{equation}
H_{f}(t)=\begin{pmatrix} \Delta_f(t) & \Omega_{f}(t) \\ \Omega_{f}(t) & -\Delta_f(t) \end{pmatrix},
\label{final_ham}
\end{equation}
where $\Delta_f(t) = [\Delta(t)-\dot{\zeta}(t)/2]/2$ and $\Omega_f = \sqrt{\Omega^2+\Omega_a^2}$. This final transformation can be realized through a simple axis rotation to the diabatic states: $\ket{\overline{\psi}_{\uparrow\uparrow}} = e^{+i\zeta(t)/2}\ket{{\psi}_{\uparrow\uparrow}}$, $\ket{\overline{\psi}_{\downarrow\uparrow}^+} = e^{-i\zeta(t)/2}\ket{{\psi}_{\downarrow\uparrow}^+}$. However that does not affect the intrinsic properties of the system.
In Fig.~(\ref{fig2}a) we show the final population of the entangled state when the dynamics is governed by $H_f(t)$ which demonstrates that even for very small values of $\Omega_0$, the final population of $\ket{{\psi}_{\downarrow\uparrow}^+}$ tends to unity. The interaction time has been scaled down to $1\%$ compared to that required in the adiabatic case. This causes violation of the adiabatic condition as $Q$ value goes up to as high as 50. However, the population still gets transferred along the adiabatic path as shown in Fig.~(\ref{fig2}b).
\begin{figure}
\includegraphics [height=8cm,width=8cm]{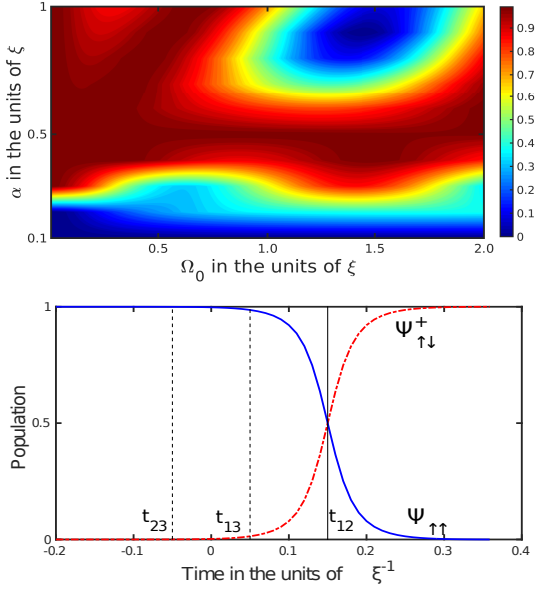}
\caption{\label{fig2} (Color online) (a) Final population of the entangled state $\ket{\psi_{\downarrow\uparrow}^+}$ against $\Omega_0$ and $\alpha$ using Transitionless driving algorithm with the parameters chosen as $\omega = \xi,\;T = 20\xi^{-1}$, (b) Evolution of
population when adiabatic condition is not satisfied with $Q = 50$ which requires less time to transfer the population to the entangled state.}
\end{figure}

{\it LRI based approach}-
The basic formalism for the TQD is to remove the non-adiabatic contribution from the adiabatic Hamiltonian. But in case of LRI based approach, to speed up the population transfer, a dynamical Invariant is being used which satisfies the general invariant equation, $dI(t)/dt = \partial I(t)/\partial t - i[I(t),H_I(t)]$. The interaction Hamiltonian can be expressed as a linear combination of the Pauli matrices: $H_I(t) = \frac{\Omega(t)}{\sqrt{2}}{\sigma}_x + \frac{\Delta(t)}{2}{\sigma}_z $. Therefore $H_I(t)$ possesses $SU(2)$ symmetry (as Pauli matrices satisfy the Lie algebra) and hence a Hermitian dynamical invariant can be constructed. We write this invariant in the following way \cite{10,18}:
\begin{equation}
I(t) =\frac{\kappa_0}{2}(\sin \gamma \cos \beta {\sigma}_x - \sin \gamma \sin \beta {\sigma}_y + \cos \gamma {\sigma}_z)
\label{inv}
\end{equation}
Here $\kappa_0$ is an arbitrary constant, which has the dimension of frequency. The eigenstates of $I(t)$ with eigenvalues $\lambda=\pm 1$, are as follows:
\begin{subequations}
\begin{align}
\ket{n_+(t)} &= \cos({\frac{\gamma(t)}{2}})e^{i\beta} \ket{\psi_{\uparrow \uparrow}} + \sin({\frac{\gamma(t)}{2}}) \ket{\psi_{\downarrow\uparrow}^+}\label{n+}\\
\ket{n_-(t)} &= \sin({\frac{\gamma(t)}{2}}) \ket{\psi_{\uparrow \uparrow}} + \cos({\frac{\gamma(t)}{2}}) e^{-i\beta} \ket{\psi_{\downarrow\uparrow}^+}\label{n-}
\end{align}
\label{eigenvect}
\end{subequations}
The parameters $\gamma$ and $\beta$ are both time dependent and they characterize $I(t)$ . Upon substituting $I(t)$ in the invariant equation, we derive the following conditions that are
required for $I(t)$ to be dynamical invariant:
\begin{subequations}
\begin{align}
\dot{\gamma}(t) &= \sqrt{2} \Omega_{LR} \sin \beta(t)\label{gam}\\
(\Delta_{LR}(t)+\dot{\beta}(t)) \sin\gamma(t) &= \sqrt{2}\Omega_{LR}(t) \cos \gamma(t) \cos \beta(t)\; \; 
\end{align}
\label{invariance}
\end{subequations}
In invariant-based approach we generally construct the fields $\Omega_{LR}(t)$ and $\Delta_{LR}(t)$ from $\gamma(t)$ and $\beta(t)$. Eq.~(\ref{invariance}) predicts the nature of dependence of $\Omega_{LR}(t)$ and $\Delta_{LR}(t)$ on $\gamma(t)$ and $\beta(t)$. Another notable thing is that adiabatic states in Eq.~(\ref{adia_basis}) are related to $\ket{n_{\pm}(t)}$ (Eq.~(\ref{eigenvect})) via Lewis-Riesenfeld phase $\eta_{\pm}(t)$ by the relation, $\ket{\phi_{\pm}(t)} = e^{i\eta_{\pm}(t)}\ket{n_{\pm}(t)}$. Thus these two sets of eigenstates do not coincide or in other words $H_I(t)$ does not commute with $I(t)$. To inverse engineer this system we design $I(t)$ through the parameters $\gamma(t)$ and $\beta(t)$ with specific boundary conditions so that it commutes with $H_I(t)$ at least at the start and at the end of the evolution i.e., $[H_I(t_i),I(t_i)] = 0 = [H_I(t_f),I(t_f)]$. In this way both $H_I(t)$ and $I(t)$ share same eigenstates at the boundaries. To achieve such a scenario the following conditions should be satisfied:
\begin{subequations}
\begin{align}
\Omega_{LR}(t)\sin \gamma(t) & \sin \beta(t)\bigr\rvert_{t=t_i,t_f} = 0\\
\sqrt{2}\Omega_{LR}(t) \cos \gamma(t)\bigr\rvert_{t=t_i,t_f}&\nonumber \\
-\Delta_{LR}(t) \sin \gamma(t) & e^{\pm i\beta(t)} ] \bigr\rvert_{t=t_i,t_f} = 0
\end{align}
\label{bound}
\end{subequations}
\begin{figure}
\includegraphics[height=4cm,width=9cm]{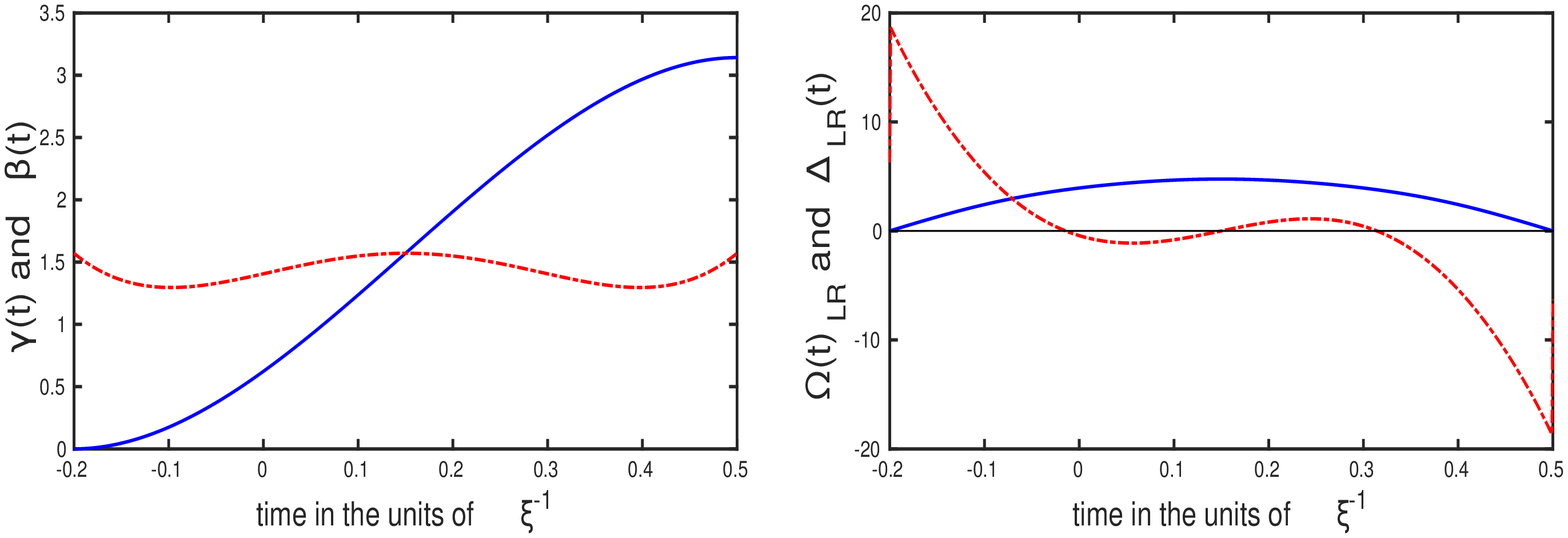}
\caption{\label{fig3} (Color online) (a) Polynomials $\beta(t)$ (dashed red) and $\gamma(t)$ (solid blue) derived from the boundary conditions by using polynomial approach with $\gamma(t) = \sum_{j=0}^4 g_j t^j$ and $\beta(t) = \sum_{j=0}^5 b_j t^j$, (b) Forms of the designed external field $\Omega_{LR}(t)$ (solid blue) and $\Delta_{LR}(t)$ (dashed red).}
\end{figure}
We set our boundary conditions by assuming $\ket{n_{\pm}(t)}$ as our instantaneous eigenstate along which the evolution will take place. As we are driving our system from $\ket{\psi_{\uparrow \uparrow}}$ to the target state $\ket{\psi_{\downarrow \uparrow}^+}$ (with or without a phase factor), Eq.~(\ref{n+}) leads us to the following conditions:
$$\gamma(t_i)=0, \; \; \gamma(t_f)=\pi$$ 
To satisfy Eq.~(\ref{bound}) following conditions are also needed:
$$\Omega_{LR}(t_i) = 0, \; \; \Omega_{LR}(t_f) = 0$$
These choices are quite in agreement with our choice of the external field for adiabatic evolution. $\Omega_{LR}(t)$ also depends on $\dot{\gamma}(t)$ and to complete the boundary 
conditions for $\gamma(t)$,using Eq.~(\ref{gam}), we can write
$$\dot{\gamma}(t_i)=0, \; \; \dot{\gamma}(t_f)=0$$
%
\begin{figure}[h!]
\includegraphics[height=5cm,width=9cm]{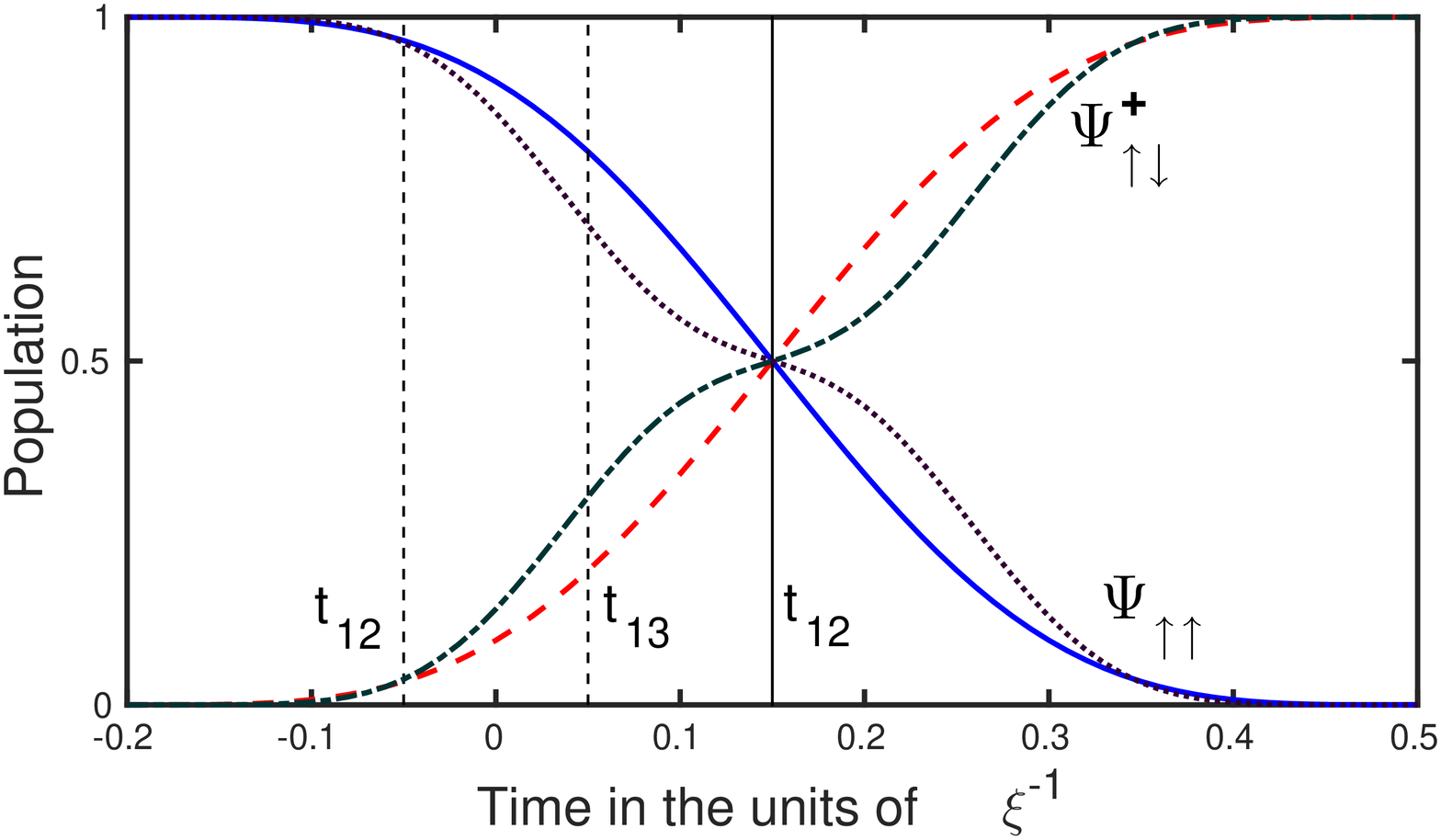}
\caption{\label{fig4} (Color online) Evolution of population for $H(t)$ and $I(t)$ using the functions $\beta(t),\;\gamma(t),\;\Omega_{LR}(t),\;\Delta_{LR}(t)$. $H(t)$ follows an
adiabatic-like path (dashed red and solid blue) while $I(t)$ does not (dotted purple and dot-dashed green).}
\end{figure}
On the other hand, the choice of $\beta(t)$ is also important for both $\Omega_{LR}(t)$ and $\Delta_{LR}(t)$. However its choice does not affect the final population of the states, but in order to keep $\Omega_{LR}(t)$ finite and minimum we restrict ourselves to the
following choices:
$$\beta(t_i)=\frac{\pi}{2}, \; \beta(t_f)=\frac{\pi}{2}, \; \dot{\beta}(t_i)=\frac{-\pi}{t_f}, \; \dot{\beta}(t_f)=\frac{\pi}{t_f}$$
The form of $\dot{\beta}(t)$ is decisive in case of designing $\Delta_{LR}(t)$. For the adiabatic case we chose it to be linear. Similarly, here we took the initial and the final boundary values of $\dot{\beta}(t)$ in such a way that it tends to show a linear nature near $t_{12}$. 
To interpolate $\gamma(t)$ and $\beta(t)$ through the intermediate temporal
points, we follow the polynomial ansatz. Two polynomials $\gamma(t) = \sum_{j=0}^4 g_j t^j$
and $\beta(t) = \Sigma_{j=0}^5 b_j t^j$ are subjected to the
above stated boundary conditions. In order to keep $\Omega_{LR}(t)$ centered at $t_{12}$ we choose the following additional 
conditions:
$${\beta}(t_{12}) = \frac{\pi}{2}, \;\; \ddot{\beta}(t_{12}) = 0, \;\; \ddot{\gamma}(t_{12}) = 0$$
With such choices $\Delta_{LR}(t)$ could be made cross through the diabatic crossing to replicate the adiabatic system. Fig.~(\ref{fig3}a) shows the nature of time dependent functions $\gamma(t)$ and $\beta(t)$, which are determined by using the boundary conditions. In Fig.~(\ref{fig3}b) we depict the functions $\Delta_{LR}(t)$ and $\Omega_{LR}(t)$ that are derived using Eq.~(\ref{invariance}). To determine the evolution of population, as shown in Fig.~(\ref{fig4}), we put $\Delta_{LR}(t)$ and $\Omega_{LR}(t)$ in the interaction Hamiltonian and also $\gamma(t)$ and $\beta(t)$ into the Invariant to solve the Schr{\"o}dinger equation numerically. The dynamics of the Hamiltonian follows adiabatic path while the invariant does not, however the end results are same for both the cases.

\begin{figure}
\includegraphics[height=5cm,width=9cm]{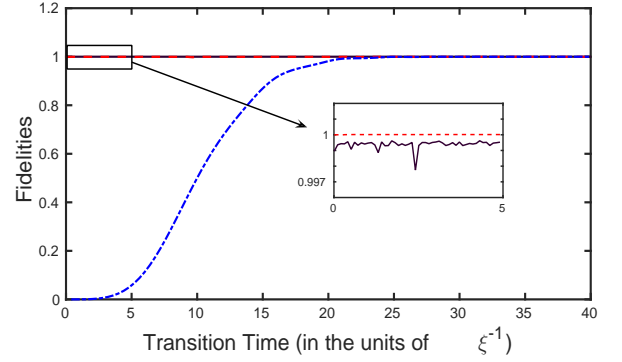}
\caption{\label{fig5}(Color online) Fidelity $\mathscr{F}$ in terms of total transition time for
three different approaches, adiabatic (dash-dotted blue), TQD approach (solid purple) and LRI based approach (dashed red).}
\end{figure}
%
{\it Fidelity}- 
Both the methods discussed above shows high fidelity in terms of population switching. Here we define fidelity as
\begin{equation}
\mathscr{F} = \lvert \braket{\psi_{\downarrow\uparrow}^+|\phi_{\pm}(t_f)}\rvert^2
\end{equation}
In Fig.~(\ref{fig5}) we have shown fidelities with respect to the
variation in total transition time. Under the adiabatic regime
fidelity shows a slow growth, which imply larger time
required for population transfer. However in case of other
two methods fidelity is close to unity regardless of the total
transition time that is again well beyond the adiabatic
condition. The TQD approach shows a value up to $\mathscr{F} = 0.999$,  while in case of the LRI based approach fidelity goes even more close to unity.

{ \it Conclusions}- In conclusion we have applied a set of effective and highly fidel shortcuts to Adiabaticity methods in a system of two spin particles coupled by an exchange
interaction and an external magnetic field. Adiabatic evolution can be used to produce a final entangled state by removing the level crossings between the diabatic states.
However it takes a large amount of time and a sufficiently strong external field to achieve perfect population transfer to the entangled state. Introduction of Transitionless
quantum driving in this system overcomes these issues. Also, this method is  robust against strong variation of $\overline{B}_z(t)$. Moreover, the design of the additional pulse entirely depends on the adiabatic parameters itself. On the other hand, in Invariant based approach we have freedom to design $\Omega_{LR}(t)$ and $\Delta_{LR}(t)$ only within a set of relevant boundary conditions (we chose polynomial approach in our
case). Although here we have presented the shortcut methods in a simple system of two spin $\frac{1}{2}$ particles, this study could further be extended to any two qubit systems. Moreover, in principle, a generalized scheme for entanglement formation with any arbitrary number of spins may also be possible.

K. Paul gratefully acknowledges a research fellowship from MHRD, Govt. of India.

\end{document}